\documentstyle[prl,aps,epsfig,amssymb,preprint]{revtex}
\begin{document}

\title{Selfsimilar Domain Growth, Localized Structures and \\
Labyrinthine Patterns in Vectorial Kerr Resonators}
\author{R. Gallego, M. San Miguel and R. Toral}
\address{Instituto Mediterr\'aneo de Estudios Avanzados, IMEDEA \cite{www}
(CSIC-UIB),\\ Campus Universitat Illes Balears, E-07071 Palma de
Mallorca, Spain.}

\date{\today}

\maketitle

\begin{abstract}
We study domain growth in a nonlinear optical system useful to
explore different scenarios that might occur in systems which do
not relax to thermodynamic equilibrium. Domains correspond to
equivalent states of different circular polarization of light. We
describe three dynamical regimes: a coarsening regime in which
dynamical scaling holds with a growth law dictated by curvature
effects, a regime in which localized structures form, and a
regime in which polarization domain walls are modulationally
unstable and the system freezes in a labyrinthine pattern.
\end{abstract}


The problem of the growth of spatial domains of different phases has been
thoroughly studied in the context of the dynamics of phase transitions: a
system is placed in an unstable state and one considers its relaxation to the
state of thermodynamic equilibrium \cite{GuntonSM}. This process is dominated
by the motion of domain walls and other defects. It is in this context that
seminal ideas of selfsimilar evolution and dynamical scaling were introduced
for nonequilibrium processes. Asymptotic domain growth laws, with their
underlying physical mechanisms, have been well established, and dynamical
scaling has been generally demonstrated. A growth law $R(t)\sim t^{1/2}$ holds
for dynamics with no conservation law and domains made of equivalent phases.
This law follows from the minimization of surface energy, and it has been shown
to be robust against the appearance of point defects in systems with a discrete
number of phases, three dimensional vortices or chiral domain walls
\cite{nonconserv}. Other well known growth laws \cite{GuntonSM} are $R(t)\sim
t^{1/3}$ for systems with conserved order parameter and $R(t)\sim t$ for
nonconserved dynamics with a metastable phase \cite{Lifshitz}, and also for
hydrodynamic systems in spatial dimension $d>2$ \cite{SanMiguel85}.

Domain growth in systems that do not approach a final state of thermodynamic
equilibrium is much less understood. For example, the mechanisms underlying a
growth law $R(t)\sim t^{1/5}$ in pattern forming systems in which the spatial
coupling is non purely diffusive (Swift-Hohenberg equation)\cite{Cross95} have
not been clearly identified. Other general issues that need to be considered
are the role of hamiltonian vs. dissipative dynamics \cite{Josserand}, the
effects of nonrelaxational dynamics such as one-dimensional motion of fronts
between equivalent states and spiral formation \cite{Gallego}, the emergence of
localized structures (LS) \cite{Coullet,localized}, or transverse instabilities
of domain walls leading to labyrinthine patterns \cite{Goldstein}.

Driven nonlinear optical systems offer a wealth of opportunities for the study
of pattern formation and other nonequilibrium processes in which the spatial
coupling is caused by diffraction instead of diffusion. These systems are
specially interesting because they naturally lead to the consideration of
vectorial complex fields, being the vector character associated with the
polarization of light, and also because they often support the formation of LS
\cite{optical,Taranenko,Oppo99}. Such bright light spots are being actively
considered for applications in parallel optical processing. Only very recently
domain growth has been considered in some of these systems and some growth laws
obtained from numerical simulations have been reported
\cite{Oppo99,Tlidi,Peschel}. However, clear mechanisms for the growth laws have
often not been identified, and some of these laws do not correspond
unambiguously to an asymptotic regime. In addition, the question of dynamical
scaling has, in general, not been addressed.

In this letter we consider a Kerr medium as a clear example of a nonlinear
optical system in which many of the issues and scenarios mentioned above can be
explored. We show that after switching-on a pump field, domain walls are formed
which separate regions with different polarization of light. The dynamical
evolution of these polarization domain walls leads to three different regimes.
For high pump values there is a coarsening regime for which we demonstrate
dynamical scaling with a growth law $R(t)\sim t^{1/2}$. For lower pump values
this process is contaminated by the emergence of LS formed by the collapse of
polarization domain walls to a stable bound structure. In a third regime the
system evolves into a nearly frozen labyrinthine pattern caused by a transverse
modulational instability of the polarization domain wall. These three
qualitatively different regimes have been experimentally observed in another
optical system \cite{Taranenko} and considered in the realm of Swift-Hohenberg
models \cite{Ouchi96}.

Our calculations are based on a mean field model that describes the transverse
spatio-temporal evolution of the two circularly polarized components of the
electric field complex
envelope, $E_+$ and $E_-$, in an optical cavity filled with an
isotropic self-defocusing Kerr medium and pumped with a linearly polarized
real field $E_0$ \cite{geddes,Hoyuelos}:
\begin{equation}\label{eq:model}
  \partial_tE_{\pm}=-(1-i\theta)E_{\pm}+i\nabla^2_{\bot}E_{\pm}+
  E_0 -\frac{1}{4}i\left[\vert E_{\pm}\vert^2+\beta\vert
  E_{\mp}\vert^2\right]E_{\pm}.
\end{equation}
Here $\theta$ is the cavity detuning, and $\nabla^2_{\bot}$ is the laplacian in
the transverse plane. Equations (\ref{eq:model}) are damped and driven coupled
Nonlinear Schr\"odinger equations which can be rewritten as
\begin{equation}
  \partial_tE_{\pm}=- E_{\pm}-i\frac{\delta\mathcal{F}}
    {\delta E_{\pm}^{\ast}},
\end{equation}
where ${\mathcal F}[E_+,E_-]$ is a real functional. Therefore, except for the
linear dissipative term, the dynamics can be written in Hamiltonian form. This
corresponds to a rather different dynamics than the normal relaxational
dynamics considered in systems that approach a state of thermodynamic
equilibrium. We will study different regimes for different values of the pump
$E_0$.

Equations (\ref{eq:model}) admit symmetric ($I_{s+}=I_{s-}$) and asymmetric
($I_{s+}\neq I_{s-}$) steady state homogeneous solutions, where
$I_{\pm}=|E_{\pm}|^2$. The homogeneous symmetric solution is linearly stable
for $E_0<E_{0,a}$, while the asymmetric solutions only exist for
$E_{0,a}<E_{0,b}<E_0$ and they are linearly stable for $E_{0,b}<E_{0,c}<E_0$
\cite{Hoyuelos,values}. There are two equivalent homogeneous stable solutions
for $E_{0,c}<E_0$, one in which $I_{s+}\gg I_{s-}$ and the other one, obtained
interchanging $E_+$ by $E_-$, in which $I_{s+}\ll I_{s-}$. These solutions are
elliptically polarized, but very close to being circularly polarized, because
one of the two circularly polarized components dominates. For simplicity we
will call them the right and left circularly polarized solutions. If the pump
field $E_0$ is switched-on from $E_0=0$ to a value $E_0>E_{0,c}$, only the mode
with zero wavenumber can initially grow from the initial condition $E_{\pm}=0$.
One then expects that either of the two equivalent homogeneous solutions will
locally grow and that domains separated by polarization walls will emerge. This
is indeed the process that we study. We note, however, that a solution with a
stripe pattern orthogonally polarized to the pump exists for $E_0>E_{0,a}$
\cite{geddes,Hoyuelos}. This pattern solution is the one obtained by continuity
from the homogeneous symmetric solution through a Turing-like instability. We
have numerically checked that such solution remains stable for pump values
$E_0\gg E_{0,c}$, but it is not the solution approached by the physical process
just described of switching-on the pump to a value $E_0>E_{0,c}$.

\begin{figure}[h]
\centerline{\epsfig{file=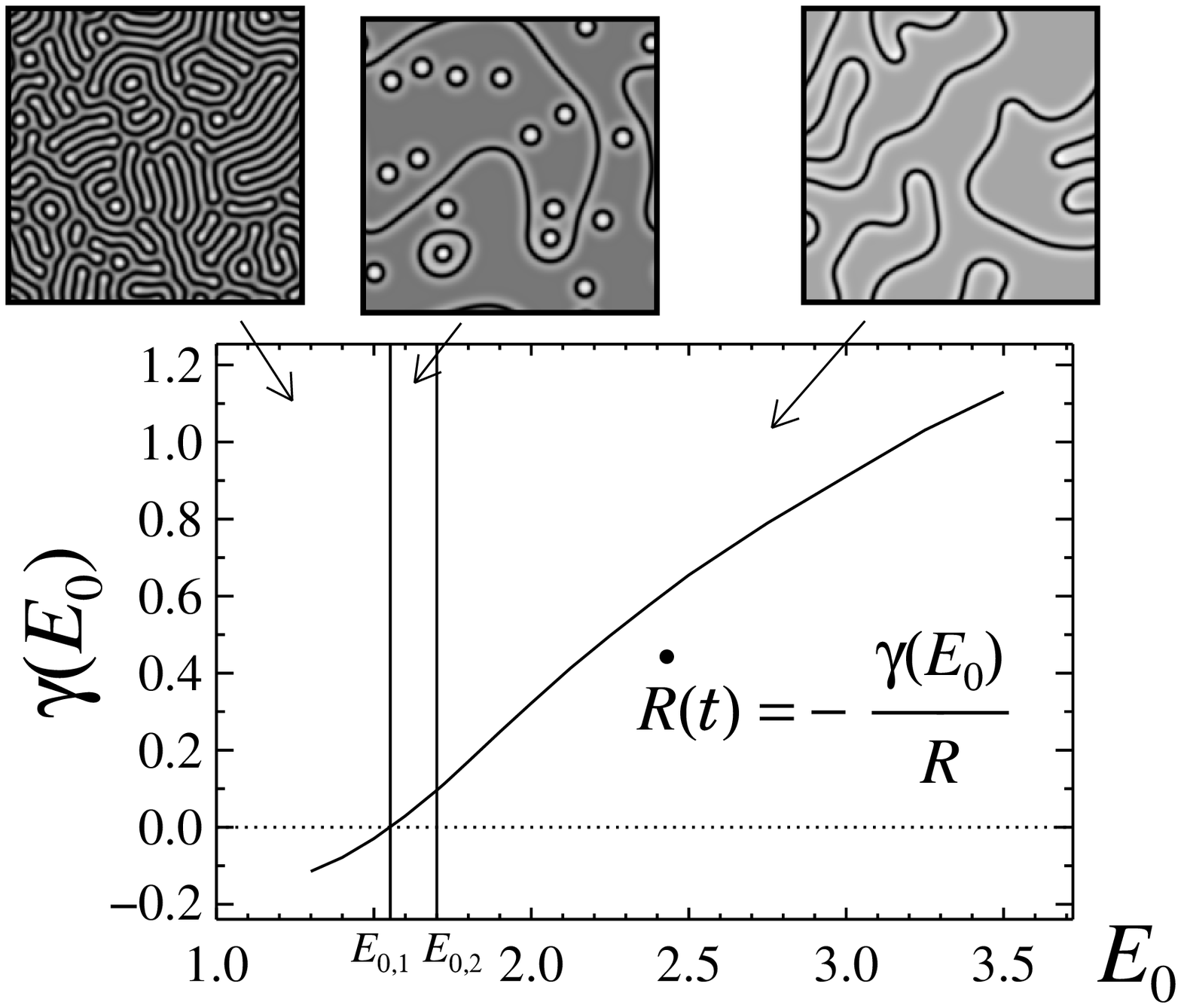, width=0.8\textwidth}}
\caption{Coefficient $\gamma(E_0)$ as defined in the text.
Snapshots of typical configurations for the total intensity $I=I_+
+ I_-$ at late times are shown for each of the three dynamical
regimes. The vertical lines identify the values of $E_{0,1}=1.552$
and $E_{0,2}=1.700$.} \label{fig:gamma}
\end{figure}

We find three different dynamical regimes for $E_0>E_{0,c}$, summarized in
fig. \ref{fig:gamma}. For $E_0>E_{0,2}$ domains grow and the system coarsens,
for $E_{0,2}>E_0>E_{0,1}$ stable LS are formed, while for $E_{0,1}>E_0>E_{0,c}$
a labyrinthine pattern emerges. These regimes are better understood by
considering the evolution of an initial isolated
polarization droplet: a circular domain of one of the solutions surrounded by
the other solution.  We find that the radius of the circular domain varies
consistently with a curvature driven front motion. The
normal front velocity $v_n$ (eikonal equation) follows a law of the form
$v_n({\mathbf r},t)=-\gamma(E_0)\kappa({\mathbf r},t)$, where $\kappa$ is the
local curvature of the domain wall and $\gamma(E_0)$ is a coefficient that
depends on the pump field amplitude. For a circular domain we get
$dR(t)/dt=-\gamma(E_0)/R(t)$. In figure \ref{fig:gamma} we show the function
$\gamma(E_0)$ as obtained from the numerical solution of eqs. (\ref{eq:model})
in a two-dimensional system
for relatively large initial droplets. Notice that $\gamma(E_0)$ changes sign
at $E_0=E_{0,1}$, which indicates a change from droplet shrinkage to droplet
growth.

\begin{figure}[h]
\centerline{\epsfig{file=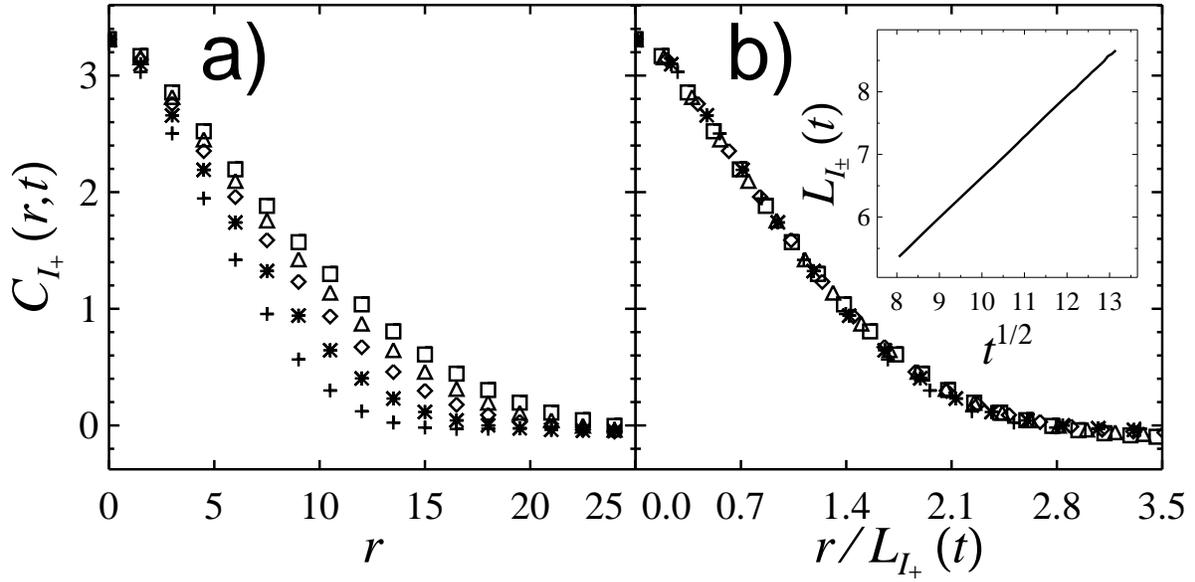,width=\textwidth}}
\caption{a) Spherical averaged correlation function for
$E_0=1.8>E_{0,2}$ at times $t=64.8$, $91.8$, $11.8$, $145.8$ and
$172.8$ and b) after scaling $r$ with the domain size
$L_{I_+}(t)$. The inset shows the domain growth law $L_{I_+}(t)
\sim t^{1/2}$.} \label{fig:scaling}
\end{figure}

We first consider the regime of domain coarsening which occurs for
$E_0>E_{0,2}$. In this regime $\gamma(E_0)>0$ and an isolated drop shrinks to
zero radius. In the general dynamics starting from random initial conditions
around $E_{\pm}=0$, sharp domain walls are initially formed and they evolve
reducing their curvature. The system approaches a final homogeneous state in
which one of the two circularly polarized solutions fills the whole system. In
order to characterize the coarsening process we have calculated the pair
correlation function of $I_+$ and $I_-$, defined as $C_{I_{\pm}}({\mathbf 
r},t)=\left\langle I_\pm({\mathbf x}+{\mathbf
r},t)I_\pm({\mathbf x},t)\right\rangle$. The average
$\left\langle\ldots\right\rangle$ is performed over the
set of points ${\mathbf x}$ (and additionally over a set of 100 different
random initial conditions). Due to the symmetry of the problem
$C_{I_+}=C_{I_-}\equiv C$. Results for the circularly averaged correlation
function $C(r,t)$ are shown in fig. \ref{fig:scaling}. The mean size $L(t)$ of
the domains is calculated as the distance at which $C(r,t)$ takes half its
value at the origin, i.e., $C(L(t),t)=\frac{1}{2}C(0,t)$. We obtain a well
defined asymptotic growth law $L(t)\sim t^{1/2}$ that follows from domain wall
motion driven by curvature effects. We have further obtained that the dynamics
is selfsimilar, i.e., that there is dynamical scaling. This is seen in fig.
\ref{fig:scaling} where we plot $C(r,t)$ before and after rescaling the spatial
coordinate of the system with the characteristic domain size $L(t)$. We observe
that curves for different times in the scaling regime collapse to the single
scaling function after rescaling. These results coincide with those obtained
for many thermodynamic systems with nonconserved order parameter
\cite{GuntonSM,nonconserv}. We note, however, that in our case the dynamics
does not follow the minimization of any obvious energy
and that surface tension is not
a proper concept for the diffractive spatial coupling considered in optical
systems.

\begin{figure}[h]
\centerline{\epsfig{file=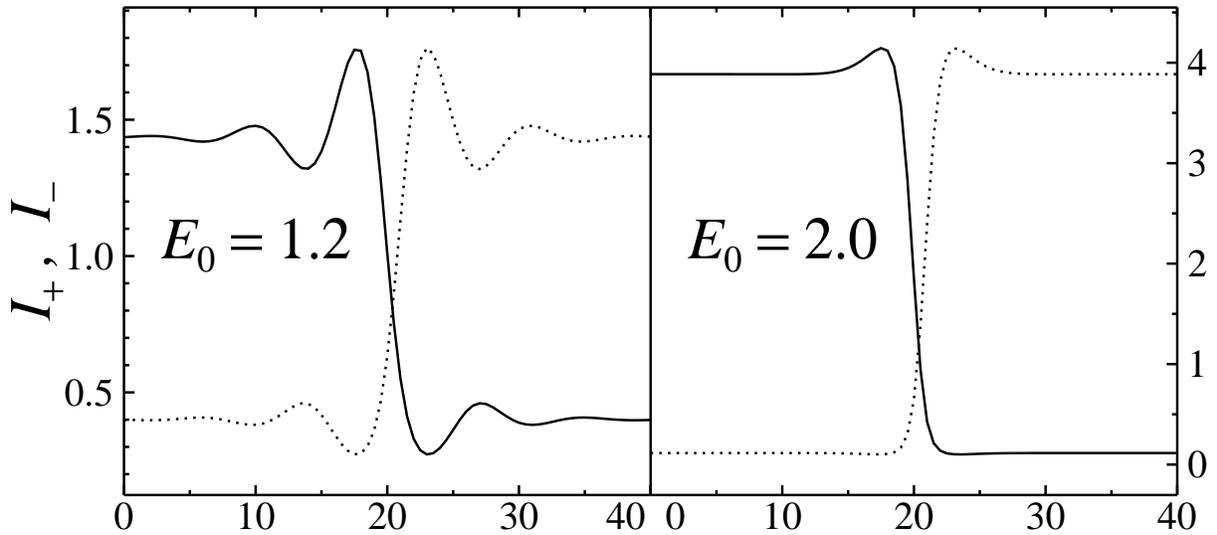,width=\textwidth}}
\vspace{0.3cm}
\caption{$d=1$ intensity profiles of the
polarization domain wall for two pump values: $E_0=1.2<E_{0,1}$,
$E_0=2>E_{0,2}$.} \label{fig:ringing}
\end{figure}

\begin{figure}[h]
\centerline{\epsfig{file=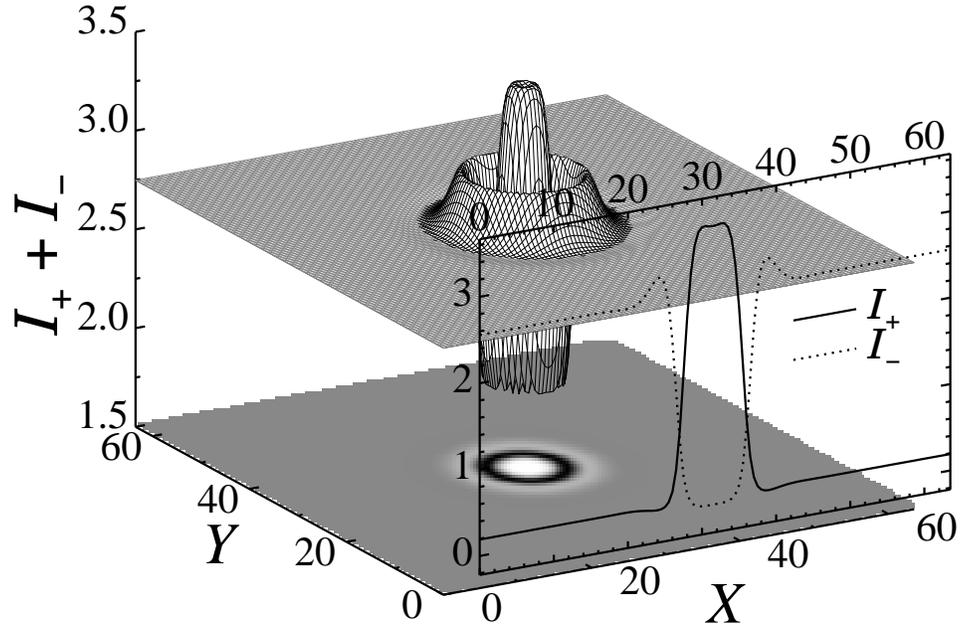, width=0.8\textwidth}}
\caption{Total field intensity ($I_+ + I_-$) of a LS
and transverse profile of $I_+$ and $I_-$ for
$E_0=1.6$.} \label{fig:LS}
\end{figure}

We next address the regime of formation of LS ($E_{0,2}>E_0>E_{0,1}$). In this
regime, as in the previous case,
 $\gamma(E_0)>0$, and a large isolated droplet initially shrinks with a
radius decreasing as $R(t) \sim t^{-1/2}$. However the shrinkage stops at a
well defined final value of the radius. Initial droplets with a smaller radius
grow to this final stable radius. In the general dynamics following the
switch-on of the pump, domain walls are initially formed. They first evolve
reducing their length as in the coarsening regime. But while in that regime a
closed loop disappears, here it collapses to a stable LS formed by a bound state
of the domain wall. The final state is composed of stretched domain walls and
LS. To understand this process is convenient to consider the form of the
polarization domain walls in a $d=1$ geometry, as shown in figure
\ref{fig:ringing}. An isolated $d=1$ domain wall is stationary. We observe that
the intensity profiles of the walls do not approach monotonically the
asymptotic value of the homogeneous state. When several domain walls are
created in the transient dynamics, they interact with each other. Since the
front profiles have oscillatory tails, the interaction between two walls can
lead to repulsive forces\cite{Coullet}. As a consequence, LS formed by bound
domain walls can be formed which stop the coarsening process. These oscillatory
tails are less important the larger is $E_0$ (see figure \ref{fig:ringing}).
However, we find that for all the values of $E_0$ which we have explored (up to
$E_0=10$), this effect is enough to stop coarsening in $d=1$: a frozen pattern
state is always dynamically reached \cite{ellipt}. What happens in our $d=2$
situation is a competition between the $d=1$ repulsive effect between walls and
the curvature effect that tends to reduce a droplet to zero radius. When the
repulsive force is large enough, it might counterbalance the shrinkage process
driven by curvature, and thus leads to the formation of a LS. This happens for
$E_{0,1}<E_0<E_{0,2}$. The mechanism is the one also discussed in
\cite{Oppo99}. These structures can be seen as a hole of $I_+$ ($I_-$) in the
background of a circularly $+$ - polarized ($-$ - polarized) state, together
with a peak of $I_-$ ($I_+$). Since the oscillatory tails are larger as $E_0$
decreases, the size of the LS decreases with $E_0$. We have found a perfect
linear dependence of the radius of the LS with $E_0$. In figure \ref{fig:LS} we
show a plot of a LS together with its transverse profile. Note that the
intensity in the LS is greater than in the surrounding background.

\begin{figure}
\centerline{\epsfig{file=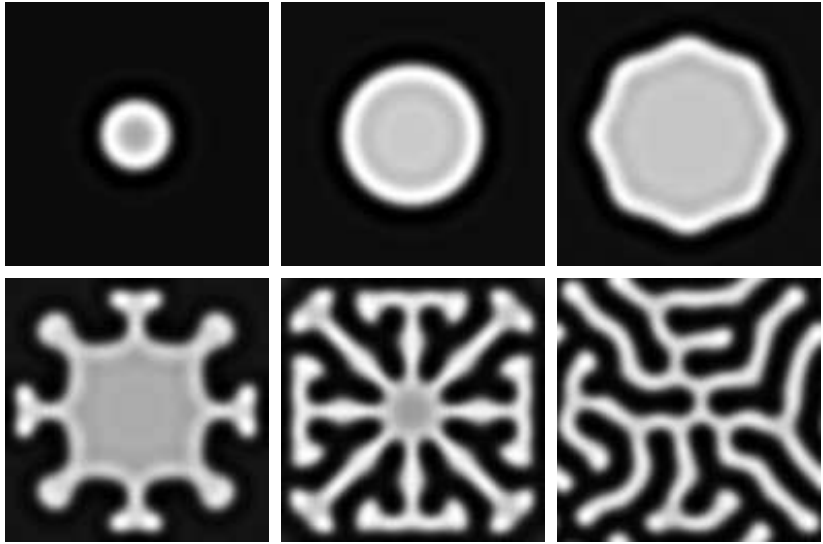, width=0.7\textwidth}}
\caption{Growth of a polarization droplet and creation of a
labyrinthine pattern for $E_0=1.3<E_{0,1}$. Snapshots at times
$t=0$, $1000$, $2100$, $2400$, $3200$ and $4900$.} \label{fig:drop}
\end{figure}

\begin{figure}
\centerline{\epsfig{file=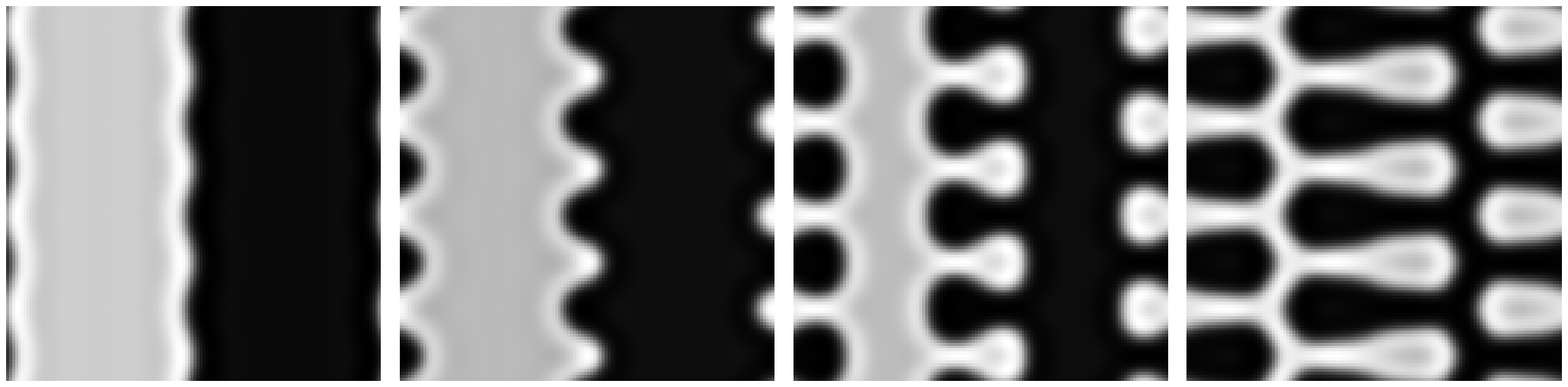, width=\textwidth}}
\caption{Transverse modulational instability for a flat domain
wall. $E_0=1.4$. Snapshots at times $t=0$, $400$, $600$, $860$.} 
\label{fig:MI}
\end{figure}

We finally discuss the regime of labyrinthine pattern formation which occurs
for $E_0<E_{0,1}$: switching-on the pump produces a very dense pattern of
domain walls that repel each other. In this regime $\gamma(E_0)<0$, and an
isolated droplet of arbitrary small size grows as $R(t) \sim t^{1/2}$. In an
infinitely large system the droplet would grow without limit, but with periodic
boundary conditions it grows until the domain wall interacts with itself.
Repulsion of the domain wall leads to a labyrinthine pattern as shown in fig.
\ref{fig:drop}. An independent way of identifying the value $E_0=E_{0,1}$,
below which labyrinthine patterns emerge, is by a linear stability analysis in
$d=2$ of the $d=1$ domain wall profile. We have numerically obtained that such
flat domain wall has a transverse modulational instability for values of the
pump amplitude for which $\gamma(E_0)<0$. We find a longwavelength instability
in which arbitrary small wavenumbers become unstable for $E_0<E_{0,1}$ (see
fig. \ref{fig:MI}). This is reminiscent of the situation described for
vectorial Second Harmonic Generation \cite{Peschel}. In physical terms, both
the droplet growth and the modulational instability indicate that the system
prefers to have the longest possible domain walls, or equivalently the largest
possible curvature. This leads to a nearly frozen state in which the
oscillatory tails of the domain walls prevent their self-crossing and in which
coarsening is suppressed. LS might form, but their natural tendency to grow is
stopped by surrounding walls.

In summary, we have described a situation in nonlinear optics in which many of
the generic issues and possible scenarios of domain growth in nonthermodynamic
systems occur. In spite of the nonrelaxational dynamics we have found a regime
of selfsimilar evolution with a growth law characteristic of curvature driven
motion. In other regimes, obtained just by changing the pump amplitude, domain
growth is contaminated by the emergence of LS or suppressed by an instability
of the domain wall that leads to a nearly frozen labyrinthine pattern. Domain
walls and LS are here associated with the polarization vectorial degree of
freedom of light.

Financial support from DGICYT (Spain, Projects PB94-1167 and
PB97-0141-C02-01) is acknowledged.
Helpful discussions with P. Colet, M. Hoyuelos and B. Malomed are also
acknowledged.


\begin{thebibliography}{99}

\bibitem[*]{www} Electronic address: http://www.imedea.uib.es/PhysDept/
%
\bibitem{GuntonSM} J.D. Gunton, M. San Miguel and P. Sahni in {\it Phase
Transitions and Critical Phenomena}, vol. 8, pp. 269-466, edited
by C. Domb and J. Lebowitz (Academic Pres, 1983); M. San Miguel,
A. Amengual, E. Hern\'andez-Garc\'{\i}a, Phase Transitions {\bf
48}, 65 (1994).

\bibitem{nonconserv}K. Kaski {\it et al.}, Phys. Rev. B {\bf 31}, 3040
(1985); M. Mondello and N. Goldenfeld, Phys. Rev. A {\bf 45},657
(1992); H. Tutu, Phys. Rev. E {\bf 56}, 5036 (1997).

\bibitem{Lifshitz} I.M. Lifshitz, Sov. Phys. JETP (1962).

\bibitem{SanMiguel85} M. San Miguel, M. Grant and J.D. Gunton,
Phys. Rev. A {\bf 31}, 1001 (1985).

\bibitem{Cross95} M.C. Cross and D.I. Meiron, Phys. Rev. Lett.
{\bf 75}, 2152 (1995).

\bibitem{Josserand}
C. Josserand and S. Rica, Phys. Rev. Lett. {\bf 69},
1215 (1997).

\bibitem{Gallego}
R. Gallego, M. San Miguel, and R. Toral,  Phys. Rev. E
{\bf 58}, 3125 (1998); Physica A {\bf 257}, 207
(1998).

\bibitem{Coullet}
P. Coullet, C. Elphick, and D. Repaux, Phys. Rev. Lett.
{\bf 58}, 431 (1987).

\bibitem{localized}  T. Ohta, Physica D {\bf 34}, 115 (1989); D. Battogtokh, M.
Hildebrand, K. Krischer, and A.S. Mikhailov,  Phys. Reports {\bf 288}, 435
(1997); H. Riecke, in {\it Pattern formation in continuous and coupled
systems}, edited by M. Golubitsky, D. Luss and S.H. Strogatz (Springer, New
York, 1999)

\bibitem{Goldstein}
R.E. Goldstein, D.J. Muraki, and D.M. Petrich, Phys. Rev.
E {\bf 53}, 3933 (1996).

\bibitem{optical}
M. Tlidi, P. Mandel and R. Lefever, Phys. Rev. Lett. {\bf 73}, 640
(1994); W. Firth and A.J. Scroggie, Phys. Rev. Lett. {\bf 76},
1623 (1996); C. Etrich, U. Peschel, and F. Lederer, Phys. Rev.
Lett. {\bf 79}, 2454 (1997); M. Brambilla {\it et al.}, Phys. Rev.
Lett. {\bf 79}, 2042 (1997); K. Staliunas and J.V.
S\'anchez-Morcillo, Phys. Rev. A {\bf 57}, 1454 (1998).


\bibitem{Taranenko}
V.B. Taranenko, K. Staliunas, and C.O. Weiss, Phys. Rev.
Lett. {\bf 81}, 2236 (1998).

\bibitem{Oppo99}
G.L. Oppo, A.J. Scroggie, and W.J. Firth, Quantum Semiclass. Opt.
{\bf 1}, 133 (1999) [These authors have also considered dynamical scaling in an
Optical Parametric Oscillator]. 

\bibitem{Tlidi}
M. Tlidi, P. Mandel and R. Lefever, Phys. Rev. Lett. {\bf 81}, 979
(1998); M. Tlidi and P. Mandel, Europhys. Lett. {\bf 44}, 449
(1998).

\bibitem{Peschel}
U. Peschel, D. Michaelis, C. Etrich and F. Lederer, Phys.
Rev. E {\bf 58}, R2745 (1998).

\bibitem{Ouchi96} K. Ouchi and H. Fujisaka, Phys. Rev. E {\bf 54}, R3895 (1996).


\bibitem{geddes} J.B. Geddes, J.V. Moloney, E.M. Wright and W.J. Firth, Opt.
Comm. {\bf 111}, 623 (1994).

\bibitem{Hoyuelos} M. Hoyuelos, P. Colet, M. San Miguel, and D. Walgraef,
Phys. Rev. E {\bf 58}, 2292 (1998).

\bibitem{values} For the parameter values used throughout this
paper, $\theta=1$ and $\beta=7$ \cite{Hoyuelos}, one finds
$E_{0,a}=0.89$, $E_{0,b}=1.02$, $E_{0,c}=1.14$.

\bibitem{ellipt} A final homogeneous state is obtained when the
pump is not linearly polarized ($E_{0,+} \neq E_{0,-}$). An
isolated domain wall is then no longer stationary.



\end{thebibliography}
\end{document}